\documentclass[superscriptaddress,twocolumn,nofootinbib]{revtex4-2}
\usepackage{graphicx}
\usepackage{bm}
\usepackage{dsfont}
\usepackage{amsmath, amsthm, amssymb,mathrsfs}
\usepackage{setspace}
\usepackage[utf8]{inputenc}
\usepackage[toc,page]{appendix}
\usepackage{epstopdf}
\usepackage{hyperref}
\usepackage{mdframed}
\usepackage{xcolor}
\usepackage{soul}

\begin{document}

\title{The wavefunction as a true ensemble}

\author{Jonte R. Hance}
\email{jonte.hance@bristol.ac.uk}
\affiliation{Quantum Engineering Technology Laboratories, Department of Electrical and Electronic Engineering, University of Bristol, Woodland Road, Bristol, BS8 1US, UK}
\author{Sabine Hossenfelder}
\affiliation{Frankfurt Institute for Advanced Studies, Ruth-Moufang-Str. 1, D-60438 Frankfurt am Main, Germany}
\date{\today}

\begin{abstract}
In quantum mechanics, the wavefunction predicts probabilities of possible measurement outcomes, but not which individual outcome is realised in each run of an experiment. This suggests that it describes an ensemble of states with different values of a hidden variable. Here, we analyse this idea with reference to currently known theorems and experiments. We argue that the $\psi$-ontic/epistemic distinction fails to properly identify ensemble interpretations and propose a more useful definition. We then show that all local $\psi$-ensemble interpretations which reproduce quantum mechanics violate Statistical Independence. Theories with this property are commonly referred to as superdeterministic or retrocausal. Finally, we explain how this interpretation helps make sense of some otherwise puzzling phenomena in quantum mechanics, such as the delayed choice experiment, the Elitzur-Vaidman bomb detector, and the Extended Wigner's Friends Scenario.
\end{abstract}

\maketitle

\section{Introduction}

The wavefunction is weird. Its most salient feature --- that it merely predicts probabilities for measurement outcomes, rather than the outcomes themselves --- suggests that quantum mechanics is an emergent, average description of an underlying, more complicated dynamics. In this underlying theory, the time-evolution of the system would be determined and measurement outcomes could be predicted. It is because we lack information about the details of the initial state that we can only make a probabilistic prediction.  

That the wavefunction might be an emergent description of yet-to-be-discovered underlying physics is often called a hidden variable interpretation. The hidden variables are the information that is missing in quantum mechanics. This straightforward explanation for the strange properties of the wavefunction, however, has seemingly been disfavoured by Bell's and related theorems \cite{Bell1964OnEPR,Clauser1969CHSH}, and, more recently, by the Pusey-Barrett-Rudolph theorem \cite{Pusey2012Reality,Leifer_2014} (hereafter {\sc PBR}-theorem). We here want to look closer at what these theorems actually say about the wavefunction as an ensemble of different values of the hidden variables. 

The {\sc PBR}-theorem in particular builds on a classification of models put forward by Harrigan and Spekkens \cite{Harrigan2010OMF}. Their framework attempts to mathematically formalise the distinction between a wavefunction that represents reality ($\psi$-ontic) and one that represents knowledge ($\psi$-epistemic). That this is a false dichotomy was previously pointed out in \cite{Hance2021Wavefunctions}. We here want to spell out another problem with the $\psi$-ontic/epistemic framework --- it doesn't identify the hidden variables theories that Bell was trying to rule out. 

We begin with defining in what sense we take the wavefunction to be an ensemble in Section \ref{ensemble}, and explain the relation to the $\psi$-ontic/epistemic framework in Section \ref{onticepistemic}. In Section \ref{trajectories} we will discuss the most important consequence, that is the different definition of what the hidden variables describe, and how to interpret it. In Section \ref{PBRsection}, we will revisit the {\sc PBR}-theorem in this light. In Section \ref{SI} we discuss Statistical Independence, locality, and the sometimes used classification of superdeterminism and retrocausality for violations of Statistical Independence. Finally, in Section \ref{exp}, we will explain how the $\psi$-ensemble interpretation helps making sense of quantum mechanics.

\section{The \texorpdfstring{$\psi$}{psi}-ensemble interpretation}
\label{ensemble}

We will begin by explaining what exactly we mean by the wavefunction being an ensemble. We do not mean the Ensemble, or Statistical, Interpretation of quantum mechanics \cite{Ballentine1970Statistical}, according to which the wavefunction describes an ensemble of identically prepared states in identical experiments. It arguably does, but if the states are indeed identical, that's not much of an ensemble. Similarly, we do not mean Smolin's ensemble interpretation \cite{Smolin2012Ensemble}, where the ensemble considered is the ensemble of all the systems in the same quantum state in the universe. 

We mean instead that the supposedly identically prepared states in supposedly identical experiments are in fact different: they have different hidden variables and this is the reason why the measurement outcomes can be different for identical wavefunctions. We will refer to this interpretation of the wavefunction as the $\psi$-ensemble interpretation. 

In our $\psi$-ensemble interpretation, we have an underlying theory with variables that we will collectively name $\kappa$ and we will denote the state-space of all $\kappa$ with $K$. 
The $\kappa$ are generically time-dependent, $\kappa(t)$. 
These $\kappa(t)$s are not the same as Bell's hidden variables, as will become clear in a moment.

In the $\psi$-ensemble interpretation, the wavefunction in quantum mechanics emerges as an average description from the hidden variables theory, much like thermodynamics emerges from statistical mechanics. This interpretation suggests itself because of the apparent similarity of the von Neumann-Dirac equation with the Liouville equation, and also, as was pointed out in \mbox{\cite{klein2011limit}}, because the naive $\hbar \to 0$ limit of quantum mechanics results in a statistical theory.

We use the term `hidden variables' because it has become common terminology. However, we want to stress that these variables are not a priori unmeasurable. They are `hidden' merely in the sense that they do not appear in quantum mechanics. If the underlying theory was better understood, they could well become measurable one day. That is to say, the $\psi$-ensemble is an interpretation of the wavefunction in quantum mechanics, but it implies the existence of an underlying hidden variables theory. This underlying theory is better referred to as a completion or modification. This is as opposed to interpretations, such as the Transactional Interpretation \mbox{\cite{Cramer1986Transactional,kastner2013transactional}} or Decoherent Histories \mbox{\cite{gell1996quantum}}, which have the same ontological basis as standard quantum mechanics, or Modal interpretations \mbox{\cite{SEPModal}} that do not rely on an underlying completion (and do not violate Statistical Independence --- we will get to this point below).

We assume that the underlying theory is deterministic, that is, if one specifies an initial state, one can use the theory to uniquely calculate the outcome of a measurement.

$|\psi \rangle$ is, as usual, an element of a projected Hilbert space, ${\cal H}$. It fulfils the Schr\"odinger equation. We are considering a generic quantum mechanical measurement, in which $|\psi \rangle$ is prepared at time $t_{\rm p}$ and measured at time $t_{\rm m}$. The measurement is described by an orthonormal basis of pointer states that we will denote $|I \rangle$, where $I \in \{0... N-1\}$ and $N$ is the dimension of the Hilbert-space. $|I\rangle$ could be describing multiple different detectors, and might be a product-state or an entangled state when expressed in a basis of the individual detectors. The basis $|I\rangle$ implicitly contains the measurement settings at the time of measurement. 

For simplicity we will in the following assume that the basis $|I \rangle$ is time-independent. This does not mean that the detector setting cannot change. It merely means that if the detector setting before the measurement was different from the setting at the actual measurement, then it wasn't described by $|I\rangle$. 

In quantum mechanics now, the Schr\"odinger evolution of $|\psi\rangle$ will generically result in a state that is not one of the detector eigenstates at the time of measurement. In this case, we assume that $|\psi(t_{\rm m}) \rangle$ is an ensemble of different $\kappa(t_{\rm m})$, that is, a distribution $\mu(\kappa(t))$ over $K$. This means in the $\psi$-ensemble theory we have a map
\begin{equation}
    {\cal P}(K) \ni \mu(\kappa(t))\to |\psi(t) \rangle \in {\cal H}~,
\end{equation}
where ${\cal P}(K)$ is the space of all normalisable probability distributions on $K$.
That is to say, we interpret $|\psi(t_{\rm m}) \rangle$ as an ensemble because we empirically know it corresponds to different measurement outcomes. 

For each pointer state $|I\rangle$, there is a subset of hidden variables, $ \{\kappa \}_I$ that will lead to this outcome. We will refer to these subsets $\{ \kappa \}_I$ as ``clusters''. The measurement  reveals which cluster the hidden variable of the actual state belonged to. Hence, the measurement reveals some information about the hidden variable. 

The reason we have for considering this interpretation is that the wavefunction update is non-local when interpreted as a physical process. This makes it difficult to combine quantum theory with general relativity. If we hence introduce a $\psi$-ensemble interpretation, our  motivation is to develop an underlying theory which restores locality, causality, and determinism. Of course one can conceive of $\psi$-ensembles that are not local, but these (unsurprisingly) are not useful for restoring locality.
Therefore, for the rest of this paper we will only consider $\psi$-ensembles that are locally causal in Bell's sense \mbox{\cite{Bell2004Speakable,cavalcanti2012bell,norsen2011john}}.

\section{Difference between \texorpdfstring{$\psi$}{psi}-ensemble and \texorpdfstring{$\psi$}{psi}-epistemic}
\label{onticepistemic}

The introduction of the $\psi$-ensemble in the previous section sounds superficially very similar to the definition of a $\psi$-epistemic model introduced by Harrigan and Spekkens \cite{Harrigan2010OMF}. According to Harrigan and Spekkens (hereafter HS), a model is $\psi$-epistemic if one particular value of the hidden variable can correspond to more than one wavefunction. 

As a simple example, consider you have a two-state system with detector eigenstates $\left| 0 \right\rangle$ and $\left| 1 \right\rangle$. You could prepare the state so that, under the Schr\"odinger-evolution it goes to $\left| 0 \right\rangle$ with certainty. Or you could prepare it so that it goes to $(\left| 0 \right\rangle + \left| 1 \right\rangle)/\sqrt{2}$. If the wavefunction was $\psi$-epistemic, then the latter case must have some overlap with the former in the underlying hidden-variables theory because it can result in the same measurement outcome. At least that is the idea.
 
Formally, HS first define the probability $p(\kappa\vert\psi)$ that the quantum wavefunction $\psi$ is described by $\kappa$, where $\kappa \in K$ \cite{Harrigan2007ProbDistn,Maroney2012Statistical}. Next, they define the probability of getting a particular measurement outcome given $\kappa$. While their framework can be used for any measurement, here we merely need the probability $\mathcal{A}(\psi|\kappa)$ that the state $\kappa$ is measured as $\psi$. From this one obtains the support $K_\psi \subset K$ for the wavefunction $\psi$ 
\begin{equation}
    \forall~\kappa\in K_\psi,\;\mathcal{A}(\psi\vert\kappa) = 1 ~.
\end{equation}
They then take a second wavefunction, $\varphi$, with its own probability distribution $p(\kappa\vert\varphi)$, and ask what is the probability that the $\kappa$s which contribute to $\varphi$ also contribute to $\psi$. They quantify this overlap as
\begin{equation}
\Delta =\int_{K_\psi} p(\kappa\vert\varphi) d\kappa~.
\end{equation}

\begin{figure}
    \centering
    \includegraphics[width=\linewidth]{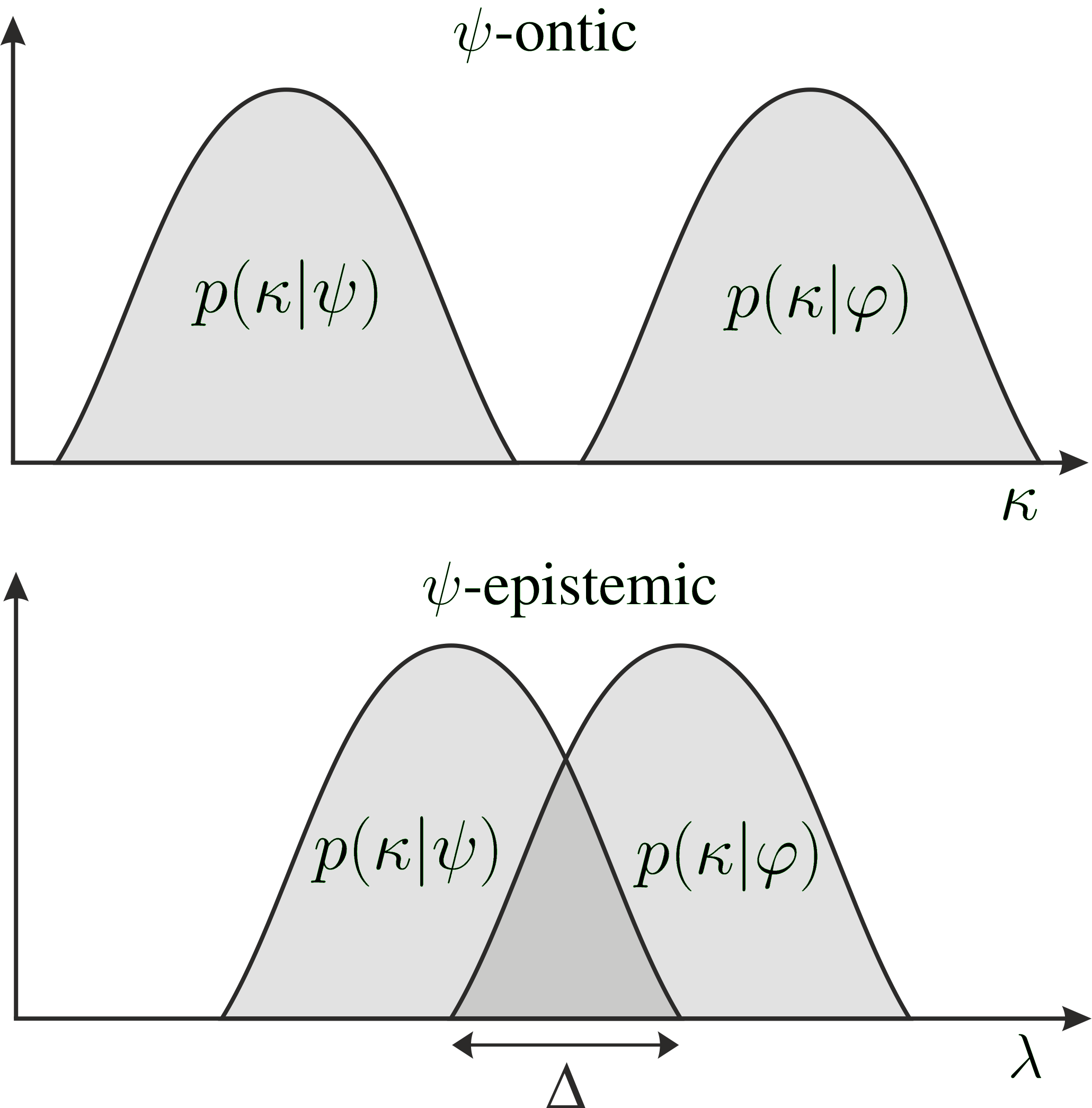}
    \caption{Harrigan and Spekkens's $\psi$-epistemic and $\psi$-ontic models of reality \cite{Harrigan2010OMF}. Wavefunctions $\psi$ and $\varphi$ each have probability distributions over state-space. In a $\psi$-epistemic model, these can overlap over a subspace. A state $\kappa$ within this overlap can then be represented both by $\psi$ and $\varphi$. However, in $\psi$-ontic models, each state can be represented by at most one wavefunction. The amount of overlap is quantified by the parameter $\Delta$. For $\psi$-ontic models, $\Delta =0$.}
    \label{fig:Graphs}
\end{figure}

If $\Delta>0$, HS interpret the wavefunctions in such a model as epistemic states, because the same underlying $\kappa$ could contribute to both $\psi$ and $\varphi$. Models with non-zero overlap $\Delta$ are correspondingly called $\psi$-epistemic, whereas models without overlap are $\psi$-ontic (for illustration, see Fig.\ref{fig:Graphs}).

There are several problems with the HS $\psi$-ontic/epistemic distinction. One of them is purely linguistic. According to the HS definition, for $\psi$-ontic theories, the wavefunction in quantum mechanics itself is ontic. It seems odd to declare something ontic that can't be observed, especially given that Bohr already argued $\psi$ is far better interpreted as knowledge. Settling a hundred-year old debate by definition is not insightful. 

One might put this aside as historical nitpicking, but this linguistic issue reflects a deeper problem. If there's no observational requirement tied to calling something ontic, then every model can be made ontic just by declaring the wavefunction to be part of the hidden variables \mbox{\cite{Schlosshauer2012Implications,Leifer_2014}}. Such an easily malleable definition of `ontic' is not what one wants to base theorems on.

It is also highly confusing that, according to the HS-definition, most hidden variables models which have been proposed so far are actually $\psi$-ontic. Think back to the example with the two state system that evolves into $(|0\rangle + |1\rangle)/\sqrt{2}$ but is measured in either $|0 \rangle$ or $|1\rangle$. While the $(|0\rangle + |1\rangle)/\sqrt{2}$ state must have been made up of different underlying states (because otherwise it couldn't give rise to different outcomes) there is no reason why any one of those underlying states must have been \emph{the same} as those of a system prepared to evolve into $|0\rangle$ or $|1 \rangle$. Hence, according to the HS-definition, a $\psi$-ontic model can solve the measurement problem as well as a $\psi$-epistemic one. 

We believe our definition of a $\psi$-ensemble to be more descriptive of hidden variables models, and less ambiguous than the $\psi$-epistemology proposed by HS. 
Whether or not the state prepared to evolve into $(|0\rangle + |1\rangle)/\sqrt{2}$ had overlap with any other wavefunction, it will always be an ensemble if the underlying hidden variables theory solves the measurement problem. That is to say a $\psi$-ensemble could be either $\psi$-ontic or $\psi$-epistemic according to the HS-definition. 

\section{Hidden variables label trajectories}
\label{trajectories}

In our definition of a $\psi$-ensemble, we have been careful to point out that naturally the underlying hidden variables theory will be dynamic, hence the hidden variables will be functions of time. This is of utmost importance for locality considerations, as those usually pertain to arguments about the distribution of the hidden variables. (We will discuss this in Section \ref{SI}.)

If we evaluate the distribution of hidden variables at the time of measurement, then we can infer the measurement outcome directly from that. However, for Bell's theorem, and also for the {\sc PBR}-theorem, one normally considers the distribution of variables at the time of preparation instead. These two distributions will in general not be identical. The problem is, if one merely takes the distribution at the time of preparation, this will \emph{not} constitute the hidden variables of Bell's theorem. 

To see this, note that since we assume the underlying theory is deterministic, we can always take the distribution at the time of preparation, $t_{\rm p}$, and evolve it forward with whatever is the transport function of the theory. Assuming time-translation invariance, let us denote the transport function as $T(t_1-t_0,\cdot)$ where the free slot is for the hidden variables. It has the property that
\begin{equation}
\kappa(t_1) = T(t_1-t_0,\kappa(t_0))~,~T(0,\cdot) \equiv {\rm Id}~.
\end{equation}
Now arguably the entire information that is necessary to determine the outcome at $t_{\rm m}$ is \emph{both} the initial distribution at preparation $\kappa(t_{\rm p})$ and the transport function $T(t_{\rm m}-t_{\rm p},\cdot)$.

Bell, therefore, in the derivation of his inequality, correctly defines the hidden variables as the `full specification' of the information necessary to predict the outcome \mbox{\cite{Bell2004Speakable}}. In his own words, the ``values of these variables together with the state vector determine precisely the results of individual measurements'' \mbox{\cite{bell1966problem}}. However, an initial state without an evolution law does not determine a final state.

Therefore, if we call Bell's hidden variables (as usual) $\lambda$, then $\lambda(t_0) = (\kappa (t_0),T(t_{\rm m}-t_0,\kappa(t_0))$ and one can now treat the hidden variables as time-independent. The additional information in the transport function becomes redundant in case one defines the distribution of the hidden variables at the time of measurement already ($t_0 = t_{\rm m}$) but normally one chooses $t_{\rm p}<t_0<t_{\rm m}$. 

\begin{figure}
    \centering
    \includegraphics[width=0.7\linewidth]{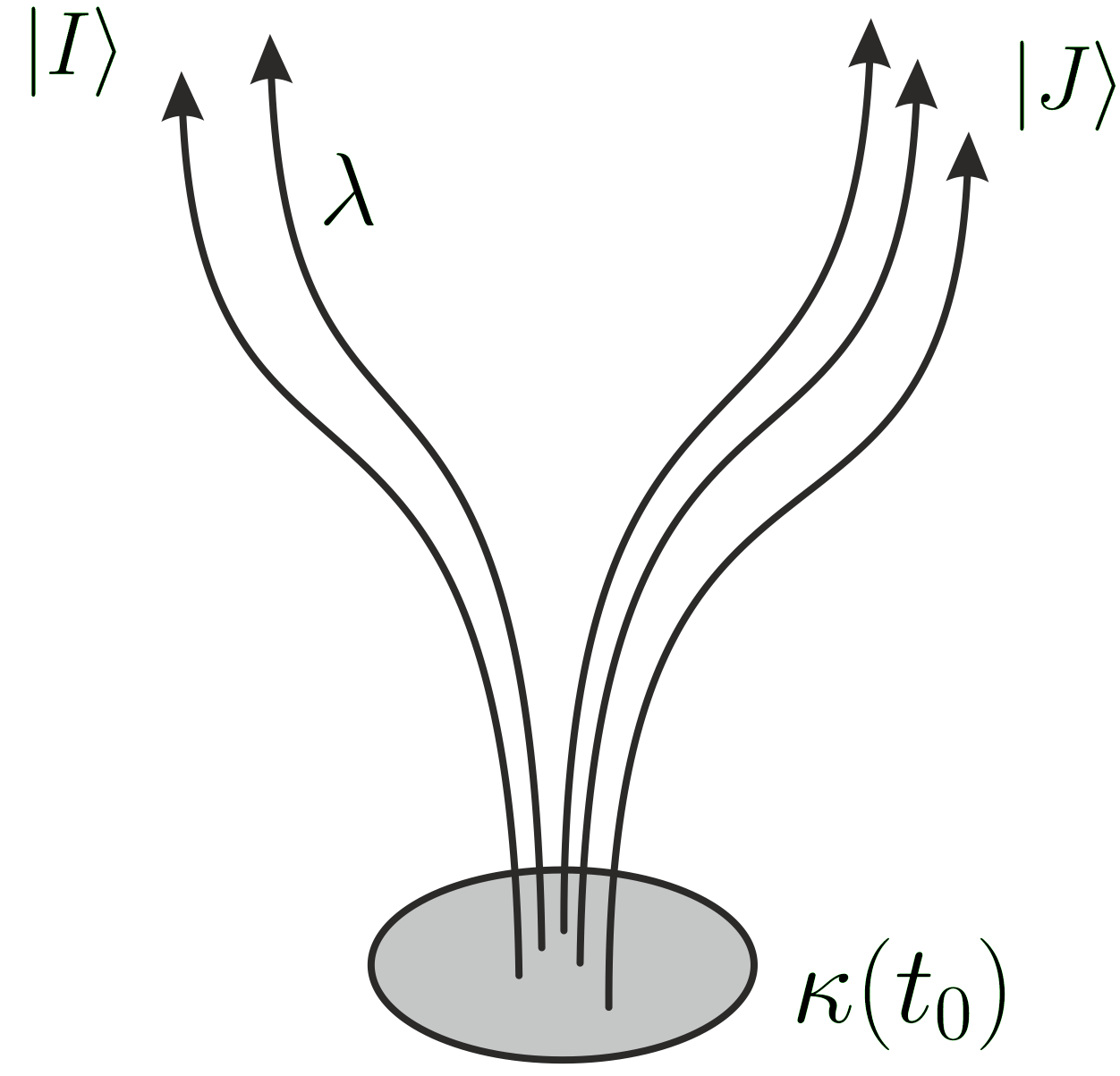}
    \caption{Sketch to illustrate the relation between the hidden variables $\kappa(t)$ which define the state, and the hidden variables $\lambda$ which define the entire trajectory. $|I \rangle$ and $|J\rangle$ depict two different detector eigenstates. The trajectories that go to the eigenstate $|I \rangle$ belong to cluster $\{ \kappa \}_I$, and those which go to $|J\rangle$ belong to cluster  $\{ \kappa \}_J$, respectively.}
    \label{fig:kappalambda}
\end{figure}

Pictorially this means that Bell's hidden variables (the $\lambda$) can be interpreted as labels for histories (in state space), whereas the HS hidden variables (that we called $\kappa$) define a state (see Fig. \ref{fig:kappalambda}). For this reason, assuming that Bell's hidden variables belong to a particular moment in the history of the (entire) state makes no sense. We will comment in Section \ref{SI} on what this means for locality requirements that one normally uses for the distribution of the hidden variables. 

A particularly illustrative example for hidden variables that define trajectories is the model proposed in \cite{Palmer2009ISP,Palmer2020Discretization}. In this case, the hidden variables simply are the detector eigenstates towards which the prepared state evolves. Given that we cannot measure the state before we measure the state, this is the most minimal assumption that one can make. 

Following the terminology introduced in \mbox{\cite{cavalcanti2012bell}}, we note that a deterministic model may not be predictable, in the sense that a model may contain variables from which one {\emph{could}} calculate measurement outcomes when one had them, and yet one cannot find out the values of the hidden variables to actually do that. Bohmian mechanics is an example of such a deterministic, yet unpredictable mode. The $\psi$-ensembles we consider here are all deterministic. They may or may not also be predictable. In the model laid out in \mbox{\cite{Palmer2009ISP,Palmer2020Discretization}} the hidden variables are incalculable with also makes the model unpredictable. 

A few words are in order here about the sense in which the $\psi$-ensemble is local. In the $\psi$-ensemble, the hidden variables are \mbox{\emph{in general}} not localised in the sense that they do not belong to some compact region of space because they actually describe trajectories. The initial values $\kappa(t_0)$ are in general not localised either because they are not uniquely defined. Suppose, for example, you have a distribution of variables in space and instead replace them with their Fourier-transformation. That would contain the same information but one may be localised, the other not. However, the initial states are localised if one considers experiments in which the prepared state is localised and is detected in localised detectors, which is for all practical purposes always the case, and certainly in Bell-type experiments. It is for this reason that the $\psi$-ensembles can fulfil the definition of local causality, as Bell intended.

We should add here that it is quite possibly the case that the type of model we consider, in which the hidden variables actually describe trajectories, is not what Bell meant when he conceived of his definition of local causality. But regardless of what Bell may have meant, such models can formally fulfil the condition of local causality.

\section{The PBR theorem revisited}
\label{PBRsection}

Let us then look at what happens when one confuses these two sets of hidden variables.
The {\sc PBR}-theorem \cite{Pusey2012Reality}, in a nutshell, shows that models which are $\psi$-epistemic according to the HS-definition are incompatible with observations. This superficially sounds similar to Bell's theorem which takes on hidden variables theories, but the {\sc PBR}-theorem works entirely differently. For this reason, one cannot draw conclusions from it about $\psi$-ensemble theories. It is not our intention here to criticise the PBR theorem which has instilled much fruitful discussion and helped sharpen terminology. We merely want to clarify what conclusions can be drawn from it.

The above discussion on $\psi$-ensembles is relevant as the hidden variables in the HS-definition merely define the state. They do not also specify the time-evolution of the underlying hidden variable system. If one hence does not define the state at the time of measurement, then the hidden variables in the HS-definition do not determine the measurement outcome. How could they if they do not contain information about the evolution law? 
Now one may say that a definition is just that, a definition. However, for the {\sc PBR}-theorem one treats the hidden variables \emph{as if} they did determine the measurement outcome, which implicitly assumes that the transport function contains no further information.

If however the transport function contains necessary information to determine the outcome, then one runs into the following problem. For the {\sc PBR}-theorem one considers the distribution of hidden variables at preparation in the HS-definition. However, those don't tell us what the outcome is, so no conclusions about whether the theory is viable or not can be derived. If one instead replaces the variables in the {\sc PBR}-theorem with the variables which actually determine the outcome --- i.e., those which contain information about the transport function --- then one needs an additional assumption. This assumption is that the complete variables are not correlated with the detector settings. That is, one needs the assumption that Bell called Statistical Independence. This is because if Statistical Independence is not fulfilled, then one of the assumptions for the {\sc PBR}-theorem, the Preparation Independence Postulate ({\sc PIP}) is not fulfilled.

Bell defined Statistical Independence ({\sc SI}) as the absence of correlations between the hidden variables and the two detector settings which he considered in a particular experiment. However, of course one can define Statistical Independence in general as 
\begin{equation}
    \mu(\lambda | S) = \mu(\lambda)~,
\end{equation}
where $S$ are the detector settings (of whatever experiment), and $\mu$ is the probability distribution of the hidden variables. As laid out in \cite{Hance2021StatInd}, violating this condition does not necessarily require correlations between the settings and the distribution of the hidden variables over the state space; the correlations can instead be induced by the structure of the state-space. 

Statistical Independence sneaks into the {\sc PBR}-theorem because the HS-definition for a $\psi$-epistemic model assumes that the hidden variables which define the state at preparation have nothing to do with the measurement. And that sounds reasonable, until one realises that if one wants to say anything about measurement outcomes one needs to know the time-evolution of the hidden variables too. 

The {\sc PIP} now says that if we prepare two independent (spatially separate) experiments, then the distributions of the hidden variables should factorise. In the {\sc PBR} argument one first considers two experiments in isolation. Then one combines the two so prepared states --- using {\sc PIP} --- and measures them instead with a smartly chosen basis of distinguishable states. From this, one derives a contradiction: If the distribution of the hidden variables was actually the product of the distributions for the separate measurements, then one cannot reproduce the predictions of quantum mechanics. 

Needless to say, the theorem is correct for what the mathematics is concerned, but the relevant physical assumption was that the distributions of the hidden variables for the individual experiments already didn't depend on the measurement settings. So of course if one multiplies them, they still do not depend on the measurement settings and one cannot reproduce the predictions of quantum mechanics. That is, in the $\psi$-epistemic framework, we have {\sc SI}~$\Rightarrow$~{\sc PIP}\footnote{Note that it's not in general the case that $\neg$ {\sc PIP} $\Rightarrow \neg$ SI because it could be that other assumptions of the $\psi$-epistemic framework are violated rather than SI.}.  

However, the relevant assumption was the independence of the complete hidden variables (the $\lambda$s which determine the outcome) from the measurement settings, not the factorisation. In particular, the initial distribution of the variables, $\kappa(t_{\rm p})$, might well factorise at preparation. 

An example for a model in which this happens is \cite{Donadi2020SuperdetToy}. This model uses one (complex valued) hidden variable for each dimension of the Hilbert-space. These hidden variables, which correspond to $\kappa$, are uniformly distributed in the unit disk.  Hence, the model fulfils the requirements for the {\sc PBR}-theorem and yet reproduces the predictions of quantum mechanics, seemingly defying the theorem. The reason is that the dynamical law of the model depends on the detector settings, and so do the full hidden variables $\lambda$. When one uses these variables, the model violates {\sc PIP}, hence no contradiction with the theorem arises. However, since the variables then no longer describe the initial state at preparation, it's unclear why they should fulfil {\sc PIP} in the first place.

We may note that this issue does not appear in Bell's theorem, simply because Bell considers a situation where the state is prepared at one place but measured in two, whereas {\sc PBR} consider a situation where the state is prepared in two places but measured in one. In Bell's case one has no reason to assume that the distribution of the hidden variables factorises. 

For completeness, let us mention that the approach presented in \mbox{\cite{barrett2014no}} to rule out (HS) $\psi$-epistemic models works by constructing sets of incompatible states and deriving properties about the overlap of probability distributions from measurements on those. This proof implicitly assumes that the probability distribution of the hidden variables does not depend on the measurement settings.

\section{Properties of \texorpdfstring{$\psi$}{psi}-ensemble theories}
\label{SI}

\subsection{Statistical Independence}

We know from Bell's theorem that any $\psi$-ensemble theory which is locally causal (in Bell's sense) and reproduces the predictions of quantum mechanics must violate Statistical Independence. Indeed, that one can reproduce quantum mechanical correlations while respecting local causality provided Statistical Independence is violated is quite possibly the reason why Bell (and others after him) sought other ways to rule out Statistical Independence, notably by arguing that it requires finetuning (discussed below). 

However, one does not need Bell's theorem to see that local and causal $\psi$-ensemble models will violate Statistical Independence. We just need to note that if we want states in the hidden variables theory to evolve locally and within the light-cone into detector eigenstates, then we need information about the detector before the prepared state reaches the detector --- otherwise how do we know what state to evolve into?

For the purposes of this paper, we shall say that a hidden variables theory solves the measurement problem if it respects Bell's local causality and gives rise to detector eigenstates in each single run of an experiment with a probability distribution that agrees (to within measurement accuracy) with that of quantum mechanics. There are other ways to look at the measurement problem \mbox{\cite{maudlin1995three}}, but for the purposes of this present paper, this simplified notion that specifically applies to hidden variables theories will suffice.

\begin{figure}
    \centering
    \includegraphics[width=\linewidth]{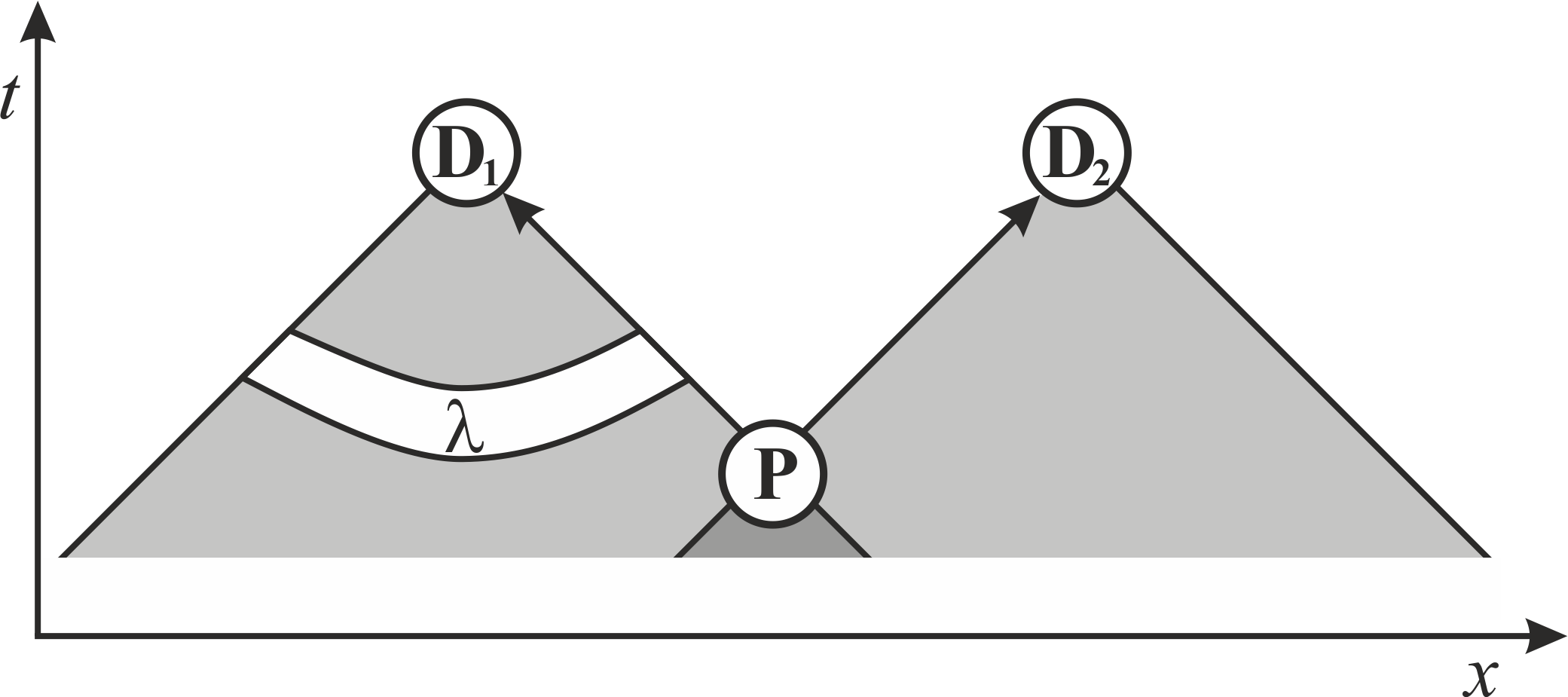}
    \caption{A diagram showing the two backwards light cones from detectors {\bf D}$_1$ and {\bf D}$_2$, and their overlap at point of preparation \textbf{P
    }.}
    \label{fig:locality}
\end{figure}

To illustrate what solutions to the measurement problem have to do with Statistical Independence, consider the generic setting of an experiment with two detectors (Fig. \ref{fig:locality}). The state is prepared at {\bf P} and measured at detectors {\bf D}$_1$ and {\bf D}$_2$. One may have in mind for example a single photon that goes through a beam splitter. The state does not have to be entangled. We want to know how it evolves in the underlying hidden variables theory. 

When the prepared state arrives at {\bf D}$_1$ it needs to know the setting at {\bf D}$_2$. Yet, it is clear that if that information wasn't available already at {\bf P}, then it can't have gotten there within the light cone. Hence, both detector settings {\bf D}$_1$ and {\bf D}$_2$ must have been either in the initial distribution of hidden variables or the evolution law. In either case, Statistical Independence is violated. 

To be sure, we have drawn an extreme case where the prepared state actually moves with the speed of light. If one took a massive particle, the last moment in which information about {\bf D}$_2$ must be available for the underlying state going to {\bf D}$_1$ could be after preparation. But that doesn't change the fact that Statistical Independence must have been violated because information about the detector setting must have been in the full specification of the hidden variables. 

It is a corollary of this --- and not the starting point --- that any $\psi$-ensemble theory that solves the measurement problem and reproduces quantum mechanics will violate the assumptions of Bell's theorem and the {\sc PBR}-theorem. It follows that experiments which show the violation of a Bell inequality or the {\sc CHSH} inequality \cite{Aspect1981BellViol}, or the agreement of the measurement proposed by {\sc PBR} with the predictions of quantum mechanics \cite{Ringbauer2015MeasRealWavefn}, cannot rule out $\psi$-ensemble theories. A different type of experiment is necessary to test such theories.  

Understanding that Statistical Independence is violated \emph{because} that is required to solve the measurement problem also explains why one need not expect such violations to occur in non-quantum experiments, like the often quoted tobacco trials to infer a correlation between smoking and cancer (see \cite{Hossenfelder2020Rethinking} and references therein). No preparation of such an experiment will ever result in a superposition of measurement outcomes. Even if one used a quantum experiment to, say, sample mice into two groups (one exposed to tobacco smoke, the other not), then the quantum superposition would be destroyed before one even wrote down the outcome of the quantum experiment. But Statistical Independence is only violated between the preparation of an experiment that, under the Schr\"odinger-evolution, evolves into a superposition of detector eigenstates, and the measurement of that state. 

It must be understood here that we use the term `measurement' in the same sense as in quantum mechanics. That is, it does not necessarily require an observer or even a detector, it just requires a sufficient amount of interactions with the environment. We don't need to be specific about this because this aspect of the measurement problem was already solved by decoherence and einselection \cite{Zurek2003Decoherence}: A detector is any system that decoheres the prepared state quickly enough, and the detector eigenstates are defined as those which are stable in that process. We merely need violations of Statistical Independence to explain why we always measure a detector eigenstate and not, as decoherence predicts, a mixed state. 

It is worth stressing that the statements made in this section are not suggested interpretations of mathematical expressions, they are conclusions drawn from empirical facts. We know, as a matter of fact, that measurements yield definite results. If the results come about locally, causally, and deterministically from the prepared state, then the underlying theory must violate Statistical Independence. We also know empirically that we never observe superpositions of detector eigenstates, hence violations of Statistical Independence are negligible for macroscopic objects.

Theories which violate Statistical Independent are often further broken down into those which are superdeterministic and those that are retrocausal. The superdeterministic ones were recently further broken down into those that are and aren't supermeasured \mbox{\cite{Hance2021StatInd}}. However, the terms superdeterministic and retrocausal have been used to mean different things by different authors which has made the situation very confusing. For example, some authors assume superdeterminism is deterministic \mbox{\cite{Hossenfelder2020Rethinking}}, while others do not \mbox{\cite{Sen2020Superdet1}}. Some equate retrocausality with future input dependence \mbox{\cite{Wharton2019Reformulations}}, others do not \mbox{\cite{price1994neglected,price2012does}}. Some claim that retrocausality is just mislabelled superdeterminism \mbox{\cite{Vervoort2022Superdet}}. Yet other authors have proposed to use the terms all-at-once or atemporal \mbox{\cite{adlam2022two}}. None seem to assign much importance to the ensemble that the wavefunction should describe, which is at the centre of our definition.

We believe that the $\psi$-ensemble is a more useful classification because it captures those approaches which are candidates to solve the measurement problem(s).  

\subsection{Locality}

It is quite common to get confused here about how such a requirement --- that the evolution law depends on the measurement settings at the time of measurement --- can possibly be local. However, keep in mind that the hidden variables $\lambda$ label trajectories, not initial states, of the underlying theory. The trajectories necessarily connect the prepared state with the detectors locally and causally (if they didn't, we couldn't detect the prepared state). Hence, the requirement that the hidden variables are correlated with the detector settings at measurement can come about entirely by local propagation and by local interactions. (Again, we do not claim that such a theory necessarily has to be local. It's just that this is the case we are interested in.)

Such $\psi$-ensemble theories that violate Statistical Independence fulfil Bell's criterion of local causality because if the information about both detector settings was available already at preparation, then it is unnecessary to later add it. 

\subsection{Fine-tuning}

It is frequently argued that theories which violate Statistical Independence would need a lot of detail to specify the initial state for the detector setting at the time of measurement, and that they are thence fine-tuned ``conspiracies'' \cite{Sen2020Superdet2}. This initial state is often believed to be required long before the experiment even began, possibly as far back as the Big Bang. 

However, the idea that we need to define an initial state for the detector setting is patently absurd. In quantum mechanics, we also never define the initial condition at the Big Bang that will give rise to a specific detector setting. Indeed, we never do this in any theory. It is in practice unnecessary to ever write down this initial state, regardless of whether it is feasible in the first place. We simply make predictions for any possible measurement setting. Hence, violating Statistical Independence does not require a lot of detail; it merely requires the detector settings at the time of measurement --- an input we also use in ordinary quantum mechanics.

A second finetuning argument \cite{Wood2015FineTuning} has it that it would require specifying a lot of details for the distribution of hidden variables in order to prevent violations of Born's rule that would allow for superluminal signalling. However, leaving aside that the absolute impossibility of superluminal signalling (rather than its observed rarity) is a postulate, not an empirical fact, we can arguably always achieve that Born's rule is fulfilled simply by postulating that it is fulfilled, which one also does in quantum mechanics. 

That is, these two arguments use a double-standard: Assumptions are not called fine-tuned when used in normal quantum mechanics (detector settings at measurement and Born's rule are required to calculate probabilities) but are called fine-tuned when used in other frameworks. As already argued in \cite{Hossenfelder2020SuperdeterminismGuide}, such a notion of fine-tuning is scientifically meaningless and can at best be considered metaphysical. (Finetuning in these contexts has been discussed in more detail by Adlam \mbox{\cite{Adlam2021Finetuning}}.)

\section{Using the \texorpdfstring{$\psi$}{psi}-ensemble to make sense of quantum experiments}
\label{exp}
Of course using a $\psi$-ensemble, even if deterministic, local, and causal doesn't remove the weirdness of quantum mechanics. It remains in the property that for a spatially distributed detector, a state in the underlying theory that goes to one particular eigenstate must have had information about that part of the detector to which it \emph{didn't} go. Again, this requirement can be formulated purely in local terms (if the state would have gone to the other part of the detector, it would have interacted with it), but this is where the famed quantum weirdness goes. If the underlying hidden variables theory is local and causal and deterministic, then the evolution which the actual state takes --- and which we ultimately observe --- takes into account what would have happened if the state had gone elsewhere. We will now explain how this nicely explains some otherwise very unintuitive quantum behaviour.

\subsection{The Double-Slit Experiment}

To warm up, we consider the familiar double-slit experiment. The odd thing about this experiment is that the behaviour of the quantum particle changes depending on whether we know which slit the particle goes through. If we don't know, the particle seems to go through both slits and interferes with itself, like we expect for a wave. But if we measure which slit the particle goes through, the interference pattern vanishes. 

This seemingly strange behaviour makes total sense from the $\psi$-ensemble perspective because the trajectory of the state in the underlying hidden variables theory depends on what is being measured. 

\subsection{Wheeler's Delayed-Choice Experiment}

The delayed choice-experiment \cite{Wheeler1978Delayed} comes in a large number of variations, but in the simplest version, one changes the variable that is being measured after the prepared state has begun its propagation. In normal quantum mechanics, this is even more difficult to understand than the double slit experiment because it seems that the prepared state must somehow `change its mind' as the setting is changed. 

However, from the $\psi$-ensemble perspective this is clearly not what happened. This is because the trajectories depend on the detector setting \emph{at the time of measurement}. What the settings were before that or how often they were changed is irrelevant.

\subsection{The Elitzur-Vaidman Bomb Detector}

The Elitzur-Vaidman bomb detector allows one to detect whether a bomb is live (would blow up when it detects a photon) without triggering the bomb \cite{Elitzur1993Bomb}. In this experiment, the bomb acts as a detector. Whether the bomb is live or a dud therefore constitutes two different detector settings. From the $\psi$-ensemble perspective, it is hence rather unsurprising that the case in which the bomb does not blow up contains information about whether the bomb is live, because this is just a type of detector setting.

\subsection{Counterfactual Computation/Communication}

In counterfactual computation \cite{Hosten2006CFComp} one infers the result of a computer query without interacting with the computer. This experiment is a variant of the Elitzur-Vaidman experiment and can be explained the same way. The computer acts as a detector depending on what the outcome of the calculation is. In the $\psi$-ensemble interpretation, the particle which one measures contains information about the setting of the detectors which it wasn't detected in, which allows one to infer the result of the computation.

The same logic also applies to Salih et al's counterfactual communication device and related protocols \cite{Salih2013Protocol,Salih2018Laws,Hance2021Quantum}, and so protocols for imaging \cite{Hance2021CFGI} and for quantum information transfer derived from it \cite{Salih2020DetTele,Salih2021EFQubit}.

\subsection{The Extended Wigner's Friend Scenario}

The previous examples all work more or less the same way. The Extended Wigner's Friend experiment \cite{Frauchiger2018Wigner} is a more recent proposal that beautifully highlights the problems one runs into when rejecting hidden variables theories, but works entirely differently. 

The Extended Wigner's Friend Scenario is a Bell-type test with two entangled particles, one measured by Charlie and one by Debbie (the friends) in spatially separated laboratories. Each of the two observers is observed by a super-observer, Alice and Bob respectively. Charlie and Debbie measure an entangled state and Alice and Bob then further measure correlations between Alice's and Bob's measurement outcomes. It turns out that, given suitable measurement settings, the observers cannot all agree on what the measurement results are. 

The issue with this experiment is a fuzzy definition of what constitutes a measurement. As already pointed out in \cite{relano2018decoherence,zukowski2021physics}, one can interpret the friends' measurements as either having resulted in a definite outcome, or as still being in an entangled quantum state. The resulting contradiction does not occur if one has a theory in which a measurement outcome is an unambiguous result of the time-evolution. 

In a $\psi$-ensemble theory, the friends either make a measurement and violations of Statistical Independence are for all practical purposes destroyed. Then the super-observers can no longer measure quantum correlations between the friends. Or the friends really just make a transformation on a quantum state, rather than a measurement, in which case the super-observers can well observe further quantum effects. 

\section{Conclusion}
We have argued here that the $\psi$-ontic/epistemic distinction is not descriptive of hidden variables models that can restore locality and causality by giving a physical meaning to the wavefunction update. Instead, they can better be described as $\psi$-ensemble theories. If they are local, causal, deterministic and reproduce the predictions of quantum mechanics, then such theories violate the assumptions of both Bell's theorem and the {\sc PBR}-theorem. We have shown that taking this possibility seriously can help making sense of some otherwise strange quantum phenomena.

\section{Acknowledgements}
We thank James Ladyman, Sophie Inman, Tim Palmer and John Rarity for useful discussions. 
SH acknowledges support by the Deutsche Forschungsgemeinschaft (DFG, German Research Foundation) under grant number HO 2601/8-1. JRH is supported by the University of York's EPSRC DTP grant EP/R513386/1, and the EPSRC Quantum Communications Hub (funded by the EPSRC grant EP/M013472/1).

\end{document}